\documentclass{PoS}

\usepackage{epsf}
\usepackage{amssymb}
\usepackage{amsmath}
\usepackage{amsfonts}
\usepackage{cite}
\usepackage[small]{caption2}

\usepackage[T1]{fontenc}
\usepackage{psfrag,epsfig,graphicx,graphics}

\newcommand{\be}{\begin{equation}}
\newcommand{\beq}{\begin{equation}}
\newcommand{\ee}{\end{equation}}
\newcommand{\eq}{\end{equation}}
\newcommand{\eeq}{\end{equation}}

\newcommand{\bea}{\begin{eqnarray}}
\newcommand{\eea}{\end{eqnarray}}

\def\slashchar#1{\setbox0=\hbox{$#1$}
   \dimen0=\wd0
   \setbox1=\hbox{/} \dimen1=\wd1
   \ifdim\dimen0>\dimen1
      \rlap{\hbox to \dimen0{\hfil/\hfil}}
      #1
   \else
      \rlap{\hbox to \dimen1{\hfil$#1$\hfil}}
      /gdatdafinal2.tex
   \fi}

\newcommand{\veck}{{\bf k}}

\newcommand{\veckone}{{\bf k}_1}
\newcommand{\vecktwo}{{\bf k}_2}
\newcommand{\veckj}{{\bf k}_{J}}

\newcommand{\veckjone}{{\bf k}_{J,1}}
\newcommand{\veckjtwo}{{\bf k}_{J,2}}

\newcommand{\deins}[1]{{\rm d}#1\,}
\newcommand{\dzwei}[1]{{\rm d}^2#1\,}
\newcommand{\dk}{\dzwei{\veck}}

\newcommand{\dkone}{\dzwei{\veckone}}
\newcommand{\dktwo}{\dzwei{\vecktwo}}

\newcommand{\dsigma}{\deins{\sigma}}
\newcommand{\dsigmahat}{\deins{{\hat\sigma}_{\rm{ab}}}}
\newcommand{\dnu}{\deins{\nu}}

\newcommand{\dx}{\deins{x}}
\newcommand{\dxone}{\deins{x_1}}
\newcommand{\dxtwo}{\deins{x_2}}
\newcommand{\dyjetone}{\deins{y_{J,1}}}
\newcommand{\dyjettwo}{\deins{y_{J,2}}}
\newcommand{\dphij}{\deins{\phi_{J}}}
\newcommand{\dphijone}{\deins{\phi_{J,1}}}
\newcommand{\dphijtwo}{\deins{\phi_{J,2}}}
\newcommand{\dtwojets}{{\rm d}|\veckjone|\,{\rm d}|\veckjtwo|\,\dyjetone \dyjettwo}
\newcommand{\shat}{{\hat s}}

\newcommand{\chihat}{{\omega}}

\title{The first complete NLL BFKL study of Mueller Navelet jets at LHC}

\ShortTitle{The first complete NLL BFKL study of Mueller Navelet jets at LHC}

\author{D.~Colferai\\
        Dipartimento di Fisica, Universit{\`a} di Firenze, Italy\\
        INFN, Florence, Italy\\
        Email: \email{colferai@fi.infn.it}}

\author{F.~Schwennsen\\
        Deusches Elektronen-Synchrotron DESY, Hamburg, Germany \\
        Email: \email{florian.schwennsen@desy.de}}

\author{L.~Szymanowski\\
        Soltan Institute for Nuclear Studies, Warsaw, Poland {\em \&} \\
        CPHT, {\'E}cole Polytechnique, CNRS, 91128 Palaiseau Cedex, France\\ 
        Email: \email{lech.szymanowski@fuw.edu.pl}}

\author{\speaker{S.~Wallon}
\\
LPT, Universit{\'e} Paris-Sud, CNRS, 91405 Orsay, France \ {\em \&} \\
UPMC Univ. Paris 06, facult\'e de physique, 4 place Jussieu, 75252 Paris Cedex  05, France\\
        E-mail: \email{wallon@th.u-psud.fr}}

\abstract{We report on the first next-to-leading BFKL study of the cross section and azimuthal decorrellation of Mueller Navelet jets. This includes next-to-leading corrections to the Green's function as well as next-to-leading corrections to the Mueller Navelet vertices. The obtained results for standard observables proposed for studies of Mueller Navelet jets show that both sources of corrections are of equal and big importance for final magnitude and final behavior of observables, in particular for the LHC kinematics investigated here in detail. The astonishing conclusion of our analysis is that the observables obtained within the complete next-lo-leading order BFKL framework of the present contribution are quite similar to the same observables obtained within next-to-leading logarithm DGLAP type treatment. The only noticeable difference is the ratio the azimuthal angular moments  $\langle \cos 2\varphi\rangle / \langle \cos \varphi\rangle$ which still differs in both treatments.
}
\FullConference{The 2011 Europhysics Conference on High Energy Physics-HEP 2011,\\
		July 21-27, 2011\\
		Grenoble, Rh\^one-Alpes France}

\begin{document}

\section{Introduction}
\label{Sec_Int}

The high energy regime of QCD is one of the key questions of particle physics.
In the semi-hard regime of a scattering process in which $s \gg -t$,  logarithms of the type $[\alpha_s\ln(s/|t|)]^n$ have to be resummed, giving the 
%
%
%
leading logarithmic (LL) Balitsky-Fadin-Kuraev-Lipatov (BFKL) \cite{BFKL_LL} Pomeron contribution to the gluon Green's function describing  the $t$-channel
 exchange.
To reveal this effect,
 various  tests have been proposed  in inclusive \cite{test_inclusive}, semi-inclusive  \cite{test_semi_inclusive} and exclusive processes \cite{test_exclusive}. The basic idea is to select specific observables minimizing  usual collinear logarithmic effects \`a la DGLAP \cite{DGLAP}  with respect to the BFKL one:
 the involved transverse scales should thus be of similar order of magnitude.
%
%
We here consider the Mueller Navelet jets \cite{Mueller:1986ey} in hadron-hadron colliders, defined as being separated by a large relative rapidity, while having two similar transverse energies.  
In a DGLAP scenario, an almost back-to-back emission is expected, while the allowed
 BFKL emission of partons between these two jets  leads in principle to a larger cross-section, with a reduced azimuthal correlation between them.
%
%
%
We review results of Ref.~\cite{us_MN} where both the  NLL Green function \cite{BFKL-NLL} and the NLL result for the jet vertices \cite{Bartels:vertex} are taken into account. 

\section{NLL calculation}
\label{sec:NLLcalculation}


The two hadrons collide at a center of mass energy $\sqrt{s}$ producing two very forward jets, whose transverse momenta  are labeled by Euclidean two dimensional vectors $\veckjone$ and $\veckjtwo$, while their azimuthal angles are noted as $\phi_{J,1}$ and $\phi_{J,2}$. The jet rapidities  $y_{J,1}$ and $y_{J,2}$  are related to the longitudinal momentum fractions of the jets via $x_J = \frac{|\veckj|}{\sqrt{s}}e^{y_J}$. We restrict ouselves
to fixed rapidities and transverse momenta.
%
%
%
%
For large $x_{J,1}$ and $x_{J,2}$, collinear factorization leads to
\begin{equation}
  \frac{\dsigma}{\dtwojets} = \sum_{{\rm a},{\rm b}} \int_0^1 \dxone \int_0^1 \dxtwo f_{\rm a}(x_1) f_{\rm b}(x_2) \frac{\dsigmahat}{\dtwojets},
\end{equation}
where $f_{\rm a,b}$ are the parton distribution functions~(PDFs) of a parton a (b) in the according proton.
%
The  resummation of logarithmically enhanced contributions 
are included 
through $k_T$-factorization:
\begin{equation}
  \frac{\dsigmahat}{\dtwojets} = \int \dphijone\dphijtwo\int\dkone\dktwo V_{\rm a}(-\veckone,x_1)G(\veckone,\vecktwo,\shat)V_{\rm b}(\vecktwo,x_2),\label{eq:bfklpartonic}
\end{equation}
where the BFKL Green's function $G$ depends on $\shat=x_1 x_2 s$. The jet vertices
 $V_{a,b}$ were  calculated at
 NLL order in Ref.~\cite{Bartels:vertex}.
%
Combining the PDFs with the jet vertices one writes
\begin{eqnarray}
  \frac{\dsigma}{\dtwojets} 
\!&=& \!\!\int \dphijone\dphijtwo \! \!\int \! \dkone\dktwo \Phi(\veckjone,x_{J,1},-\veckone)\,G(\veckone,\vecktwo,\shat)\,\Phi(\veckjtwo,x_{J,2},\vecktwo) \,,\nonumber
\\
\mbox{where }&& \quad \Phi(\veckjtwo,x_{J,2},\vecktwo) = \int \dxtwo f(x_2) V(\vecktwo,x_2).
\end{eqnarray}
%
%
In view of the azimuthal decorrelation we want to investigate, we define the  coefficients
\beq
  \mathcal{C}_m \equiv \!\!\!\int \!\! \dphijone\dphijtwo\cos\big(m(\phi_{J,1}-\phi_{J,2}-\pi)\big)\!\!\!\int\!\!\dkone\dktwo \Phi(\veckjone,x_{J,1},-\veckone)G(\veckone,\vecktwo,\shat)\Phi(\veckjtwo,x_{J,2},\vecktwo) , \nonumber
\eq
from which one can easily obtain the differential cross section and azimuthal decorrelation as
\begin{equation}
  \frac{\dsigma}{\dtwojets} = \mathcal{C}_0 \quad {\rm and} \quad 
  \langle\cos(m\varphi)\rangle \equiv \langle\cos\big(m(\phi_{J,1}-\phi_{J,2}-\pi)\big)\rangle = \frac{\mathcal{C}_m}{\mathcal{C}_0} .
\end{equation}
The guiding principle of the calculation is then to use
 the LL-BFKL eigenfunctions
\begin{equation}
  E_{n,\nu}(\veckone) = \frac{1}{\pi\sqrt{2}}\left(\veckone^2\right)^{i\nu-\frac{1}{2}}e^{in\phi_1}\,,
\label{def:eigenfunction}
\end{equation}
although they
 strictly speaking do not diagonalize the NLL BFKL kernel.
In the LL approximation, 
\begin{equation}
  \mathcal{C}_m = (4-3\delta_{m,0})\int \dnu C_{m,\nu}(|\veckjone|,x_{J,1})C^*_{m,\nu}(|\veckjtwo|,x_{J,2})\left(\frac{\shat}{s_0}\right)^{\chihat(m,\nu)}\,,
\label{eq:cm2}
\end{equation}
\begin{align}
 {\rm where} \quad C_{m,\nu}(|\veckj|,x_{J})
=& \int\dphij\dk \dx f(x) V(\veck,x)E_{m,\nu}(\veck)\cos(m\phi_J) \,, \label{eq:mastercnnu}
\end{align}
and
$
  \chihat(n,\nu) = N_c\alpha_s/\pi
\chi_0\left(|n|,\frac{1}{2}+i\nu\right) ,$ with
$\chi_0(n,\gamma) = 2\Psi(1)-\Psi\left(\gamma+\frac{n}{2}\right)-\Psi\left(1-\gamma+\frac{n}{2}\right)\,.
$
%
%
%
The master formulae of the LL calculation~(\ref{eq:cm2}, \ref{eq:mastercnnu}) will also be used for the NLL calculation, the eigenvalue now turning to an operator containing a $\nu$ derivative, 
which acts on the impact 
factors and effectively leads to a contribution to the eigenvalue
 which depends on the impact factors. 



At NLL, the jet vertices are intimately dependent on the jet algorithm ~\cite{Bartels:vertex}. We here use  the cone algorithm.
At NLL,
one should also pay attention to the choice of scale $s_0$. We find the choice of
scale  $s_0 =\sqrt{s_{0,1} \, s_{0,2}}$  with 
$s_{0,1}= \frac{x_{1}^2}{x_{J,1}^2}\veckjone^2$ rather natural, since it does not depend on the momenta $\veck_{1,2}$ to be integrated out. Besides, the dependence with respect to $s_0$ of the whole amplitude can be studied,
when taking account the fact that both the NLL BFKL Green function and the vertex functions are $s_0$ dependent.
In order to study the effect of possible collinear improvement \cite{resummed},
we have, in a separate study, implemented for $n=0$ the scheme 3 of the first paper of Ref.~\cite{resummed}. This is only required by the Green function since we could show by a numerical study that the 
jet vertices are free of $\gamma$ poles and thus do not call for any collinear improvement.
In practice, the use of Eqs.~(\ref{eq:cm2}, \ref{eq:mastercnnu}) leads 
to the possibility to calculate for a limited number of $m$ the coefficients $C_{m,\nu}$ as universal grids in $\nu$, instead of using a two-dimensional grid in $\veck$ space.
We use MSTW 2008 PDFs \cite{Martin:2009iq} and a two-loop strong coupling with a scale $\mu_R= \sqrt{|\veckjone|\cdot |\veckjtwo|}\,.$ 
In order 
to compare our analysis with DGLAP NLO approaches \cite{Fontannaz} obtained through the NLL-DGLAP partonic generator \textsc{Dijet} \cite{Aurenche:2008dn}, for which symmetric configurations lead to instabilities, we here display our results for $|\veckjone|=35\,{\rm GeV}$, $|\veckjtwo|=50\,{\rm GeV}$
(see Ref.~\cite{us_MN} for symmetric configurations).

\section{Results}
\label{sec:Results}

\begin{figure}[h]
  \centering
  \psfrag{varied}{}
  \psfrag{cubaerror}{}
  \psfrag{C0}{\raisebox{.1cm}{\footnotesize $\mathcal{C}_0 \left[\frac{\rm nb}{{\rm GeV}^2}\right] = \sigma$}}
  \psfrag{Y}{\footnotesize$Y$}
 \hspace{-.2cm} \includegraphics[width=4.7cm]{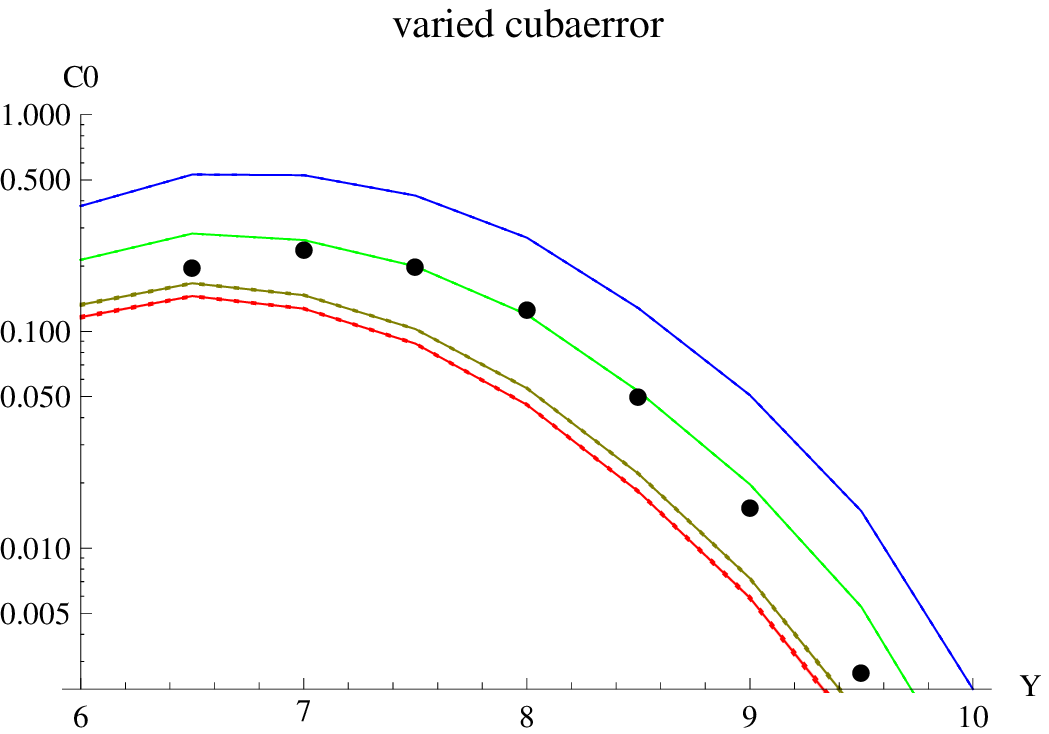} \quad
   \psfrag{C1C0}{\raisebox{.1cm}{\footnotesize $\frac{\mathcal{C}_1}{\mathcal{C}_0}=\langle \cos \varphi\rangle$}}
\includegraphics[width=4.7cm]{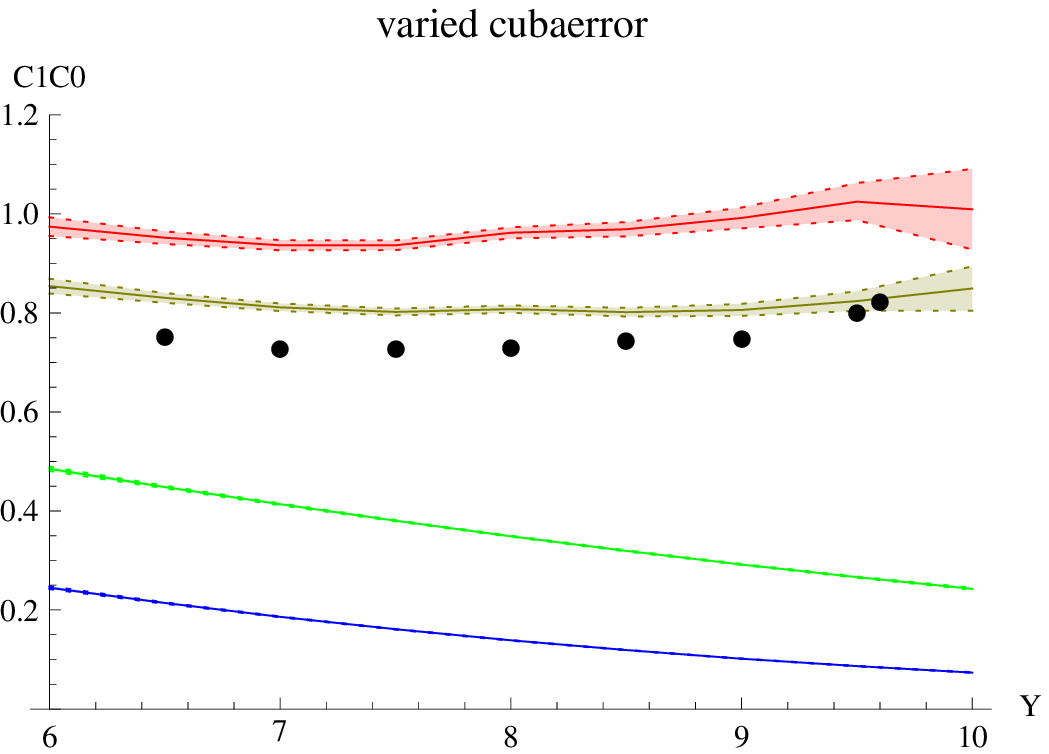}\quad 
 \psfrag{varied}{}
   \psfrag{cubaerror}{}
   \psfrag{C2C1}{\raisebox{.1cm}{\footnotesize$\frac{\mathcal{C}_2}{\mathcal{C}_1}=\langle \cos 2\varphi\rangle / \langle \cos \varphi\rangle$}}
   \psfrag{Y}{\footnotesize$Y$}
  \includegraphics[width=4.7cm]{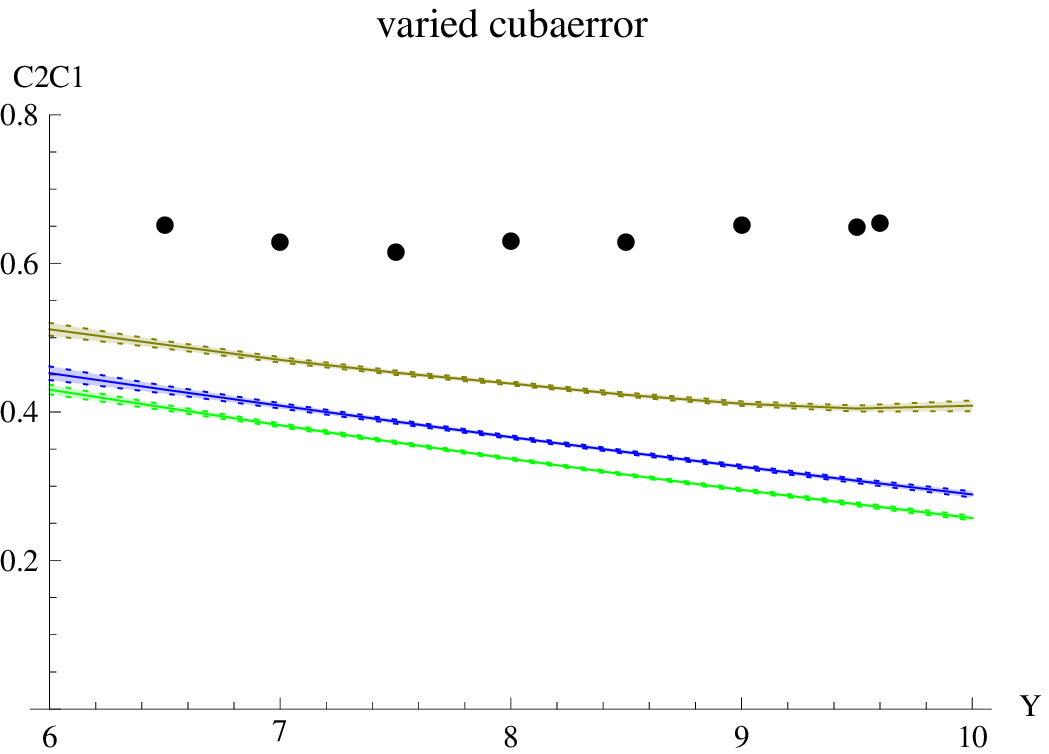}
  \caption{Differential cross section (a), azimuthal correlation $\langle \cos \varphi\rangle$ (b) and ratio $\langle \cos 2\varphi\rangle / \langle \cos \varphi\rangle$ (c) in dependence on $Y$ for $|\veckjone|=35\,{\rm GeV}$, $|\veckjtwo|=50\,{\rm GeV}$. 
The errors due to the Monte Carlo integration  are given as error bands. 
Blue: pure LL result; Brown: pure NLL result; Green: combination of LL vertices with the collinear improved NLL Green's function; Red: full NLL vertices with the collinear improved NLL Green's function.
  Dots show the results of Ref.~\cite{Fontannaz} obtained with \textsc{Dijet} \cite{Aurenche:2008dn}.}
  \label{fig:c03550_c1c03550}
\end{figure}

Fig.~\ref{fig:c03550_c1c03550}a and \ref{fig:c03550_c1c03550}b  respectively
display the cross-section and the azimuthal correlation as a function of the relative jet rapidity $Y$, for the LHC design center of mass energy $\sqrt{s}=14\,{\rm TeV}$. 
This explicitely shows 
the dramatic effect of the NLL vertex corrections, of the same order as the one for the Green function \cite{Vera:Marquet}. In particular, the decorrelation based on our full NLL analysis is very small, similar to the one based on NLO DGLAP.
%
%
%
%
The main source of uncertainties is due to the renormalization scale $\mu_R$ and to the energy scale $\sqrt{s_0}$. \ This is particularly important for the azimuthal correlation, which, 
 when including a collinear improved Green's function, may exceed 1 for small $\mu_R=\mu_F$.
The only remaining observable for which 
a noticeable difference can be expected between BFKL and DGLAP type of treatment is the ratio 
$\langle \cos 2\varphi\rangle / \langle \cos \varphi\rangle\,.$ 

The NLL analysis presented here could be extended  to describe  forward jets production.

\vspace{.1cm}

Work supported in part by the French-Polish Scientific Collaboration Agreement 
POLONIUM, the grant ANR-06-JCJC-0084 and by a PRIN grant (MIUR, Italy).

\end{document}